\documentclass[12pt,preprint,revtex4]{aastex}
\usepackage{epstopdf,graphicx,mathtools,amsmath,amssymb}
\usepackage[normalem]{ulem}
\newcommand\WCO{\ifmmode{W_\mathrm{CO}\,}\else{$W_\mathrm{CO}$\,}\fi}
\newcommand\htwo{\ifmmode{{\rm H}_2}\else{H$_2\,$}\fi}
\newcommand{\hi}{H\,{\small I}\,}
\newcommand{\hii}{H\,{\small II}\,}
\newcommand\xco{\ifmmode{X_\mathrm{CO}\,}\else{$X_\mathrm{CO}$\,}\fi}
\newcommand\Av{\ifmmode{A_V}\else{$A_V$}\fi}
\newcommand\pers{\ifmmode{\rm s^{-1}}\else{s$^{-1}$}\fi}
\newcommand\km{\ifmmode{\rm km}\else{km}\fi}
\newcommand\kms{\km\,\pers}
\newcommand\cmtwo{\ifmmode{\rm cm^{-2}}\else{cm$^{-2}$}\fi}
\newcommand\EBV{\ifmmode{E(B-V)}\else{$E(B-V)$}\fi}
\newcommand\EBVres{\ifmmode{E(B-V)_\mathrm{res}\,}\else{$E(B-V)_\mathrm{res}$\,}\fi}
\newcommand\Avres{\ifmmode{A_{V, \mathrm{res}}\,}\else{$A_{V,\mathrm{res}}$\,}\fi}
\newcommand\rOph{$\rho$\ Oph}
\newcommand\jone{\ifmmode{J\!=\!1\!\rightarrow\!0}\else{$J\!=\!1\!\rightarrow\!0$}\fi}
\newcommand{\ee}[1]{\ifmmode{\times 10^{#1}}\else{$\times 10^{#1}\,$}\fi}
\newcommand\pdeg{\fdg}

\newcommand{\planck}{\textit{Planck}}
\newcommand{\fermi}{\textit{Fermi}}
\newcommand{\TauEB}{\ifmmode{\tau_{353}\,}\else{$\tau_{353}\,$}\fi}

\begin{document}
\title{Calibrating Column Density Tracers with Gamma-ray Observations of the $\rho$ Ophiuchi Molecular Cloud}

\author{Ryan D. Abrahams$^{1,2,3}$, Alex Teachey$^{4,5}$, and Timothy A. D. Paglione$^{1,2,3}$}
\altaffiltext{1}{Department of Physics, Graduate Center of the City
University of New York, 365 Fifth Ave., New York, NY 10016, USA}
\altaffiltext{2}{Department of Earth \& Physical Sciences, York
College, City University of New York, 94-20 Guy R. Brewer Blvd.,
Jamaica, NY 11451, USA}
\altaffiltext{3}{Department of Astrophysics, American Museum of
Natural History, Central Park West at 79th Street, New York, NY 10024, USA}
\altaffiltext{4}{Department of Physics and Astronomy, Hunter College, City University of New York, New York, NY 10065, USA}
\altaffiltext{5}{Department of Astronomy, Columbia University, 550 W 120th St., New York, NY 10027, USA}

\begin{abstract}

Diffuse gamma-ray emission from interstellar clouds results largely from cosmic ray (CR) proton collisions with ambient gas, regardless of the gas state, temperature, or dust properties of the cloud. The interstellar medium is predominantly transparent to both CRs and gamma-rays, so GeV emission is a unique probe of the total gas column density. The gamma-ray emissivity of a cloud of known column density is then a measure of the impinging CR population and may be used to map the kpc-scale CR distribution in the Galaxy. To this end, we test a number of commonly used column density tracers to evaluate their effectiveness in modeling the GeV emission from the relatively quiescent, nearby $\rho$ Ophiuchi molecular cloud. We confirm that both \hi\ and an appropriate \htwo\ tracer are required to reproduce the total gas column densities probed by diffuse gamma-ray emisison. We find that the optical depth at 353 GHz (\TauEB) from \planck\ reproduces the gamma-ray data best overall based on the test statistic across the entire region of interest, but near infrared stellar extinction also performs very well, with smaller spatial residuals in the densest parts of the cloud. 

\end{abstract}


\section{Introduction}
Diffuse gamma-ray emission from the Milky Way informs us about the highest energy processes in the Galaxy such as supernova energy deposition, cosmic ray (CR) acceleration and energy losses, and CR transport.  GeV gamma-ray emission in particular is a unique probe of hadronic CR processes.  Specifically, CR protons and nuclei  
accelerated by supernova remnants impact ambient gas, which ultimately creates secondary pion emission with a characteristic spectral bump around 1 GeV \citep{AckermannSNR}. The interstellar medium (ISM) is effectively transparent to CRs, especially above the pion production threshold of 282 MeV. Therefore, interactions between lower energy CRs and molecular clouds can cause deep heating and ionization \citep{Padovani2009}, thus affecting cloud chemistry as well as star and planet formation \citep{Glassgold2012, IndrioloMcCall2012, Grenier2015}. The higher energy interactions in turn give rise to gamma-ray emission sensitive to the total gas column density of a cloud independent of its dust properties, temperature, or gas state (\hii, \hi, or \htwo). 
 
%

Likelihood analyses of gamma-ray counts maps require modeling a variety of presumed emission sources including their spatial extents and spectral shapes. Typically, a number of gas templates based upon ISM tracers (21-cm \hi\ emission, infrared dust emission, etc.), including one to account for molecular gas poorly traced by CO, the so-called``dark gas", are compared to the gamma-ray emission. The differences between the observed { gamma-ray} counts maps and these templates often result in significant, spatially coherent residuals \citep[e.g.,][]{PlanckCham}. These residuals are distinct from the dark gas and persist despite accounting for dark gas using dust maps or other measures. Given the goal of understanding the pervading CR density, spectrum, and its spatial variation through the Galactic disk, the distribution and column density of the gas with which CRs interact must be sensitively constrained. 
Common practice is to compare the gas templates to the gamma-ray counts, then incorporate any residual counts back into the templates to recover an ad hoc cloud emissivity from the gamma-ray data. Our goal instead is to reduce systematic uncertainties introduced by this method by identiying the best fitting cloud models based on maps of ISM tracers, and thus avoid distorting information about the underlying CR population.


We chose to study the \rOph\ molecular cloud because of its proximity to Earth ($d = 120^{+4.5}_{-4.2}$ pc) \citep{Loinard2008} and relative quiescence, as well as for the many special features of the region that somewhat complicate the gamma-ray analysis.  { It has an estimated mass of 6.6\ee{3} M$_\odot$ \citep{Cambresy1999} calculated from extinction measured via star counts to 1.2\ee{4} M$_\odot$ \citep{Yang2014} calculated from from dust emission measured by the \planck\ satellite \citep{Ade2011}.} The dust properties of the \rOph\ cloud are notoriously unusual \citep{Liseau2015}, which calls into question standard scaling relations between $A_V$, $A_\lambda$, CO emission, dust emission, and gas column density.  The \rOph\ dark core harbors the star-forming region L1688, which exhibits radio and X-ray emission \citep{Dzib} potentially indicative of locally accelerated CRs.  L1688 heats the dense part of the cloud, which can cause sharply varying dust temperatures that often complicate the dust modeling and column density determinations \citep{Abergel2014}.  The molecular cloud is a known source of significant anomalous microwave emission \citep{Ade2011}, presumably an additional dust component. It also exhibits a high dynamic range in $A_V$, from widespread areas of $A_V<1$ mag, to $A_V>30$ mag near L1688. The diffuse gamma-ray emission from the cloud was studied with the \textit{COS-B} and the \textit{Compton Gamma-Ray Observatory} satellites \citep[e.g.,][]{Issa1981, Hunter1994} and with the \fermi\ \textit{Gamma-ray Space Telescope} \citep{Yang2014}.While at a relatively high latitude of $16\pdeg5$, its proximity to the Galactic center line of sight implies significant gamma-ray contributions from Galactic inverse Compton (IC) emission, \hi, gamma-ray point sources, and potentially H{\small II}.  At least two bright and highly variable radio AGN, which are difficult to model appropriately and often leave large residuals, lie behind the diffuse cloud emission near L1688.  Finally, the \rOph\ cloud lies in the immediate foreground of the northern \textit{Fermi} bubbles \citep{Su2010} as well as the Galactic center GeV excess \citep{Daylan2016}.

Assuming that gamma-rays are an unbiased and linear tracer of total gas column density, we test several gas templates based on emission and extinction measures to determine the systematic errors, biases, and appropriateness of each. Throughout this paper, we refer to the dense and extended complex of the \rOph\ molecular cloud as the ROMC to distinguish this feature from the B star $\rho$ Ophiuchi itself.

\section{Gamma-ray Data and Modeling}


We use Pass 8 data from the \textit{Fermi} LAT between August 4, 2008 and June 3, 2015. Pass 8 is the most recent data release from \fermi\ and improves upon photon detection and measurement, instrument point-spread function, and energy dispersion. Analysis of the gamma-ray data is performed with the \textit{Fermi} science tools (v10r0p5) available from the \textit{Fermi} Science Support Center \footnote{FSSC: \url{http://fermi.gsfc.nasa.gov/ssc/}}, utilizing both the P8R2\_SOURCE\_V6 and P8R2\_ULTRACLEANVETO\_V6 instrument response function\textbf{s}. When selecting the data, we consider both front and back converted photons in the ``source" and ``ultraclean veto" class\textbf{es}. These data classes differ in the strictness of background CR removal, where the ultraclean veto class has the lowest CR contamination. Both classes are analyzed separately to check for background systematics.

\begin{figure}
\centering
\includegraphics[width=\linewidth]{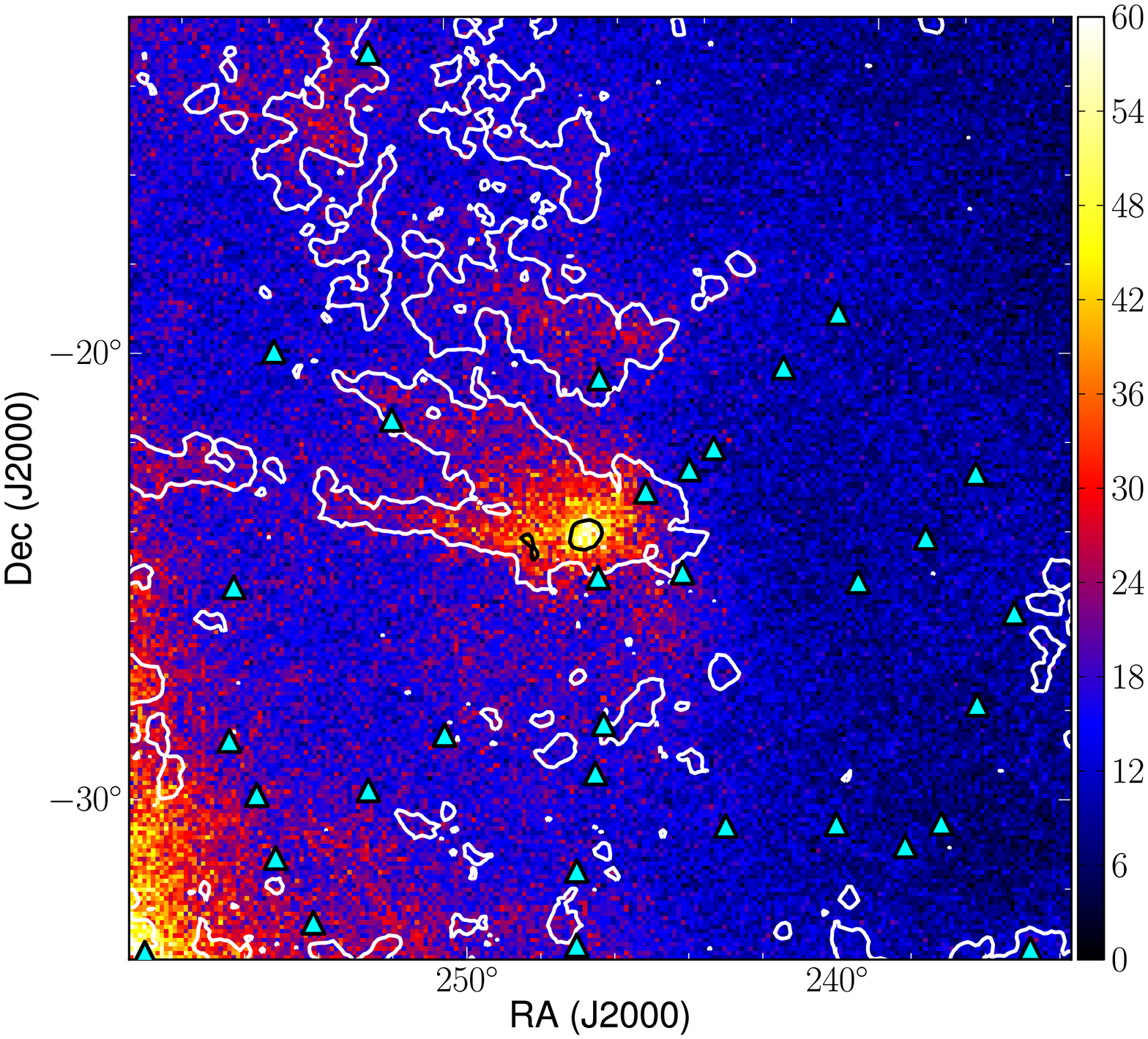}
\caption{Gamma-ray counts map from 250 MeV to 10 GeV with 0\pdeg1 pixels centered on \rOph. White and black contours represent CO integrated intensities \WCO = 2.5 and 50 K \kms, respectively. The cyan triangles are 3FGL point sources included in the model.}
\label{fig:CMAP}
\end{figure}

We downloaded data from a $15\degr$ radius around the chosen coordinates (RA, Dec = 246\pdeg4, -23\pdeg4, J2000) and between 250 MeV and 10 GeV. The data were subsequently binned between $235\pdeg8\le {\rm RA}\le257\pdeg0$ and $-34\pdeg0 \le {\rm DEC} \le -12\pdeg8$ with a spatial resolution of $0\pdeg1$ per pixel, and 30 logarithmically spaced energy bins between 250 MeV and 10 GeV.
These energies are chosen to maximize both source localization and photon statistics. Including photons above 10 GeV does not improve the significance of the detection \citep{Abrahams2015}. We excluded photons with incidence angle $>90\degr$ from the zenith and any time the spacecraft rocking angle exceeds 52\degr. These constraints removed most gamma-ray contamination coming from the Earth's limb. Figure \ref{fig:CMAP} shows the gamma-ray counts map of the region.

The models analysed broadly consisted of two nearly independent components: point sources and diffuse sources. Point sources are taken from the 3FGL \citep{Fermicat2015}, and can be seen in Figure \ref{fig:CMAP}. Within the 3FGL point source catalog, a number of sources are included that are designated as ``confused.'' In general these sources are regions of excess emission that may not be associated with any known gamma-ray source. Toward L1688 there are two such sources, 3FGL J1628.2-2431c and 3FGL J1626.2-2428c, which appear to account for the gamma-ray emission from the densest part of the ROMC, and were therefore removed.

\subsection{Diffuse Emission Templates}

Diffuse components common to every model are: the isotropic emission, Galactic IC emission, and an \hi\ map from the Galactic All-Sky Survey \citep[GASS,][]{GASS}. In addition, we test an \hii\ template \citep{Finkbeiner2003}, as there are a number of hot stars in the region.

For the IC component, we used two different models calculated with GALPROP\footnote{Source code can be found at \url{https://sourceforge.net/projects/galprop}} \citep{GALPROP98,Vladimirov2011}. One matches the older model used in \citet{Ackermann2012c}, and assumes a Galactocentric radius of 20 kpc, a maximum height between $|z| \le 4$ kpc, the \hii\ distribution from \citet{Cordes1991}, and a parameterized CR source distribution given in the GALPROP manual as ``source\_model 1". The second model assumes a larger CR halo with radius of 30 kpc, height $|z|<6$ kpc, \hii\ distribution from \citet{Gaensler2008}, and a CR source distribution following the distribution of pulsars \citep{Taylor1993}.

The isotropic emission, originating from extragalactic diffuse gamma-ray emission and misclassified cosmic rays, is modeled by the ``iso\_P8R2\_SOURCE\_V6\_v06.txt" or \\``iso\_P8R2\_ULTRACLEANVETO\_V6\_v06.txt" files provided by the Fermi Science Support Center. For the GASS \hi\ map, the relevant portion of the sky coincident with the ROI was downloaded\footnote{https://www.astro.uni-bonn.de/hisurvey/gass/}.

Finally, the \textit{Fermi} bubbles \citep{Su2010} represent an additional background source of gamma-ray emission for the region. We assume a constant counts map across the ROI and the spectrum given in \citet{AckermannBubble}.

\subsection{Molecular Gas Templates}
 The following local ISM templates were compared to capture the molecular gas component, each model incorporated one of the following maps.

\begin{enumerate}
\item the standard \textit{Fermi} Galactic diffuse model \citep{Casandjian2015FDM},

\item the 353 GHz dust optical depth, \TauEB, from the \textit{Planck} Public Data Release 1 \citep{Abergel2014},

\item a map of the $V$-band extinction, $A_V$, constructed from data taken from DSS plates and using a star counts approach \citep{DobDSS}, which we will refer to as ``Dobashi DSS,''

\item an $A_V$ map constructed from 2MASS near-infrared (NIR) photometry using the so-called ``X-percentile'' method \citep{Dob2MASS}, a modification of the Near-Infrared Color Excess (NICE) approach \citep{Lada1994}, and which we will refer to as ``Dobashi 2MASS,''

\item our own $A_V$ map using 2MASS photometry and the NICER technique as described in \citet{nicer}.
\end{enumerate}


The \fermi\ diffuse model was used for comparison for the other models. Because the \Av\ tracer presumably traces both \hi\ and \htwo, it should account for most of the gamma-ray emission from the ISM. However, dust does not homogeneously trace both diffuse and dense regions in the ISM \citep{Abergel2014}. In addition, \citet{Peek2013} showed that a combination of far infrared emission from dust and \hi\ emission best reproduced the \Av. We therefore include both \hi\ and \Av\ templates to more accurately trace the ISM.

\subsubsection{\fermi\ Galactic diffuse model}
The \fermi\ Galactic diffuse model provided by the FSSC uses the LAB survey \citep{Kalberla2005} to trace \hi, the \citet{Dame2001} CO survey to trace molecular gas, and residual color excess from \citet{SFD} to trace dark gas. The residual color excess is found by subtracting a linear combination of the integrated intensities of \hi\ and CO from the color excess.

\subsubsection{\planck\ \TauEB}
\begin{figure}
\includegraphics[width=\textwidth]{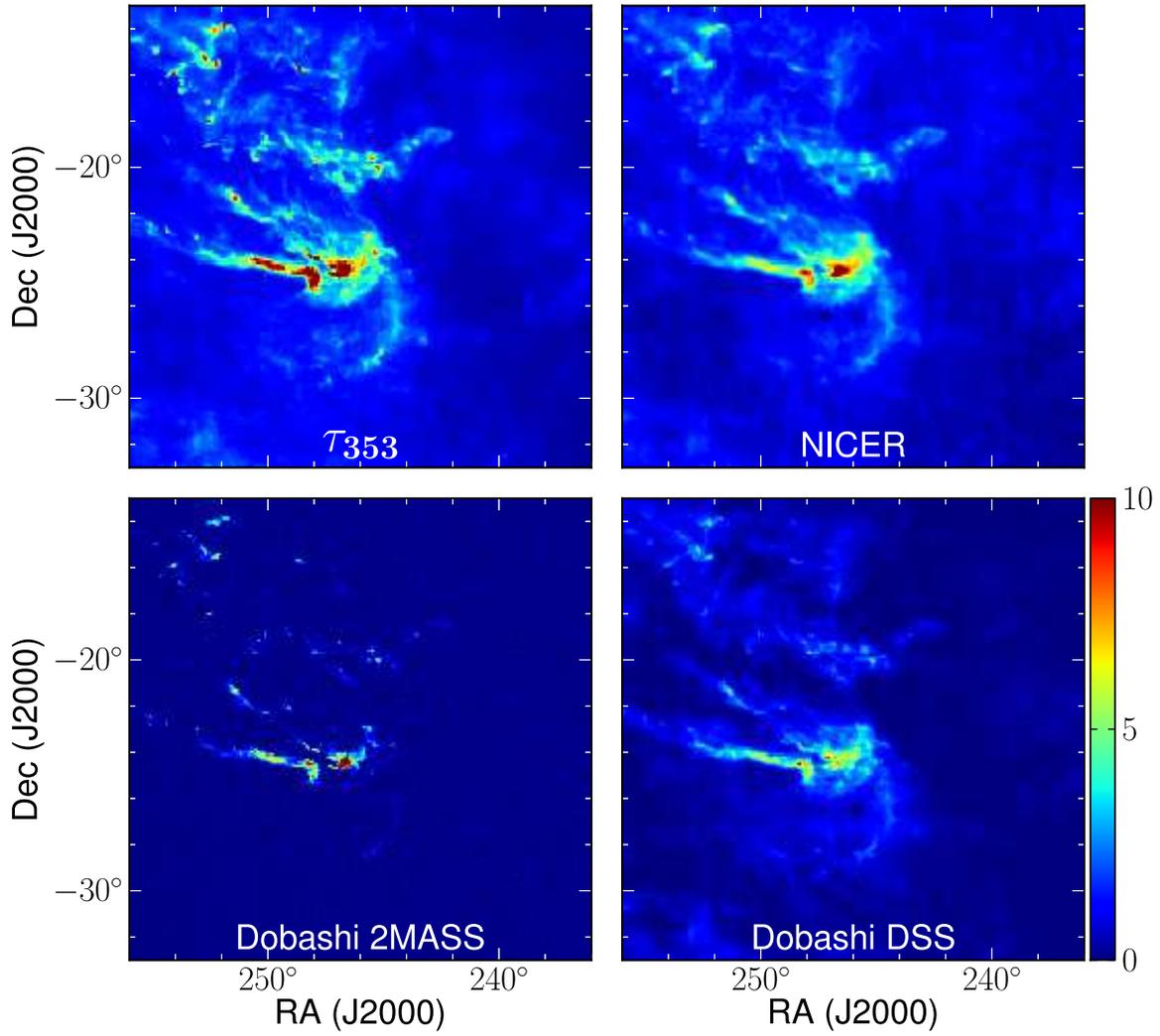}
\caption{Column density tracer templates used for four of the models. Colors represent \Av\ and range from 0 to 10 mag of extinction.}
\label{fig:temps}
\end{figure}

The all-sky optical depth map at 353 GHz from \textit{Planck}, \TauEB, was downloaded from the NASA/IPAC Infrared Science Archive (IRSA) and the ROI extracted from the HealPix version 3.10 and converted to a Cartesian projection, shown in Figure \ref{fig:temps}. For comparison to the other extinction tracers,  we converted \TauEB\ to \Av assuming $A_V = 3.1 E(B-V)$ and $E(B-V) = 1.49\ee{4}\TauEB$ \citep{Abergel2014}. 

As described in \citet{Abergel2014}, the dust opacity at 353 GHz is found by fitting a modified black body spectrum to the intensity detected by \textit{Planck} at the frequencies of the High Frequency Instrument. The optical depth is proportional to the total column density of gas:

\begin{equation}\label{eq:tau_column}
\TauEB = \sigma_{e,353} N_\mathrm{H},
\end{equation}

\noindent where $\sigma_{e,353}$ is the dust emission opacity at 353 GHz, or roughly the optical depth per H atom, and is assumed to be constant in the \planck\ \TauEB\ maps. Because \TauEB\ traces dust and not total gas directly, this $\sigma_{e,353}$ factor depends on the gas-to-dust ratio, the dust optical properties, and the proportion of gas not traced by the $N_\mathrm{H}$ proxy (dark gas). 
Dark gas, which may be ionized gas, CO-faint-\htwo, optically thick \hi, etc., serves to increase the $\sigma_{e,353}$ since there would be more gas, and thus also more dust emission, compared to $N_\mathrm{H}$ estimated from \hi\ and CO emission.  


We tested \textit{Planck} maps of $E(B-V)$ and radiance as well, but their strong dependence on dust temperature makes them quite evidently inappropriate gas models, so we mostly omit them from subsequent analysis and discussion.
%
%

\subsubsection{NICER}
Another technique for deriving an $A_V$ map using near infrared { stellar colors}, the NICER algorithm \citep{nicer} is capable of achieving a good dynamic range in $A_V$, especially in dense clouds \citep{Alves2014, nicer.Ori}. With the NICER approach, $A_V$ is calculated using $J-H$ and $H-K_s$ colors from 2MASS and by comparing the median color excess of stars in a given map pixel to an intrinsic locus in color-color space estimated from a control region assumed to have $A_V = 0$. This method is successful because of the comparatively small scatter in intrinsic NIR colors and small variations in the extinction curve \citep{Juvela2016}. Photometric errors are weighted such that more reliable colors carry more influence in the calculations. For this work we generated maps with 6\arcmin \hspace{1 pt} oversampled pixels to match the resolution of the \textit{Fermi} maps. 2MASS photometry was downloaded from the IRSA website in four pieces to avoid data clipping associated with IRSA file size limits. We filtered the data using the cc\_flg and gal\_contam markers, thereby eliminating a variety of contaminated and confused sources, and in the final map we masked the bright star Antares which produces artificially high extinction measurements close to the ROMC. The map contained $\sim$ 14 million stars after filtering. Stellar densities per pixel for this ROI range from $<10$ up to $>1000$ stars, with a median density $\sim$ 130 stars per pixel.

To calculate the intrinsic locus of an extinction-free star sample, a reference field must be selected. As both stellar density and population compositions are dependent on Galactic latitude, we tested both large and small reference fields in three low-extinction regions of the map to quantify the effect of sample size and composition on the resulting map. The differences between the resulting maps were  roughly 0.05 mag, which is an order of magnitude lower than the noise, indicating that the choice of control field sample size is largely negligible assuming the region is in fact nearly or entirely extinction-free. It is therefore suitable to select a fairly small region when observing sufficiently populated latitudes, as larger regions will necessarily sample a broader range of $A_V$. For our analysis we used a 6\arcmin $\times$ 6\arcmin \hspace{1 pt} (1 pixel) reference field sample containing $\sim$ 340 stars centered on (RA,Dec) = (253\pdeg1, -27\pdeg1). The location of the reference field was also tested and is of particular importance for the \rOph\ region, which lies in front of the large stellar density gradient of the Galactic bulge. None of the extinction-free test fields resulted in differences in inferred cloud structures, only modest, $A_V<0.4$ mag, systematic shifts in overall mean extinction over the entire ROI. Such a systematic shift in basically an extinction zero point has no effect on the renormalized templates used in the likelihood analysis, and we find no correlation with any cloud structures down to $A_V<0.05$ mag.  Maps generated with our code were compared with published maps of the ROMC \citep{Ridge2006,Lombardi2008} as well as the Corona Australis dark cloud \citep{Alves2014}, and were found to be in good agreement.


\subsubsection{Dobashi DSS}
For the Dobashi extinction maps, only the relevant fraction of the sky coincident with our ROI was downloaded\footnote{\url{http://darkclouds.u-gakugei.ac.jp/}}. \citet{DobDSS} employed a traditional starcounts approach to generate $A_V$ maps from the Digitized Sky Survey (DSS). Extinction was calculated using  

\begin{equation}
A_{\lambda}(l,b) = \frac{[\log N_0(l,b) - \log N(l,b)]}{a_{\lambda}},
\end{equation}

\noindent where $N$ is the observed stellar density, $N_0$ is the modeled background stellar density, $\lambda$ is the band of the map, and $a_{\lambda}$ is the slope of the $m_{\lambda} - log N$ diagram, the so-called Wolf Diagram \citep{Wolf1923}. As this method uses optical star counts, it is very sensitive to -- but may also saturate at -- relatively low values of $A_V$. Therefore, it is sensitive to the periphery of a dense cloud where $A_V \sim 1$ mag, but fails to properly probe the densest parts of the cloud due to low star counts.

\subsubsection{Dobashi 2MASS}
\citet{Dob2MASS} developed an alternative to the NICE method for mapping extinction, the ``X-Percentile Method'', utilizing 2MASS photometry. In this approach, the X$^{\rm th}$ percentile color excess is observed towards the cloud and compared to the X$^{\rm th}$ percentile color excess in a nearby reference field. Doing so corrects for foreground star contamination and the contamination from asymptotic giant branch stars and protostars, both of which are exceptionally red and can severely bias color excess measurements. While NIR photometry is not very sensitive to very low extinction regions, it allows more accurate values in dense clumps.





The final model may be written as

\begin{align}
I(\ell,b,E) = &\sum_i PS_i(E) \delta_i(\ell,b) + x_{\Av}q_{\rm DMN}(E) \Av(\ell,b) + x_{\rm\hi} q_{\hi}(E)N(\hi) \nonumber\\
&+ x_{\rm \hii}q_\mathrm{DNM}(E)F_{\rm\hii}(\ell,b) + x_{IC}c_{IC}(\ell,b,E) \nonumber \\
&+ x_{\rm bubbles}c_{\rm bubbles}(E) + x_{\rm ISO} c_{\rm ISO}(E).\label{eq:model}
\end{align}

\noindent Each ISM component (\Av, \hi, and \hii) is multiplied by a gamma-ray emissivity, $q_i$, and a normalization $x_i$. We follow the naming conventions for the diffuse emissivities from \citet{Casandjian2015}. The emissivities can be found in Table 1 of \citet{Casandjian2015} and were determined using Pass 7 Reprocessed data. This captures the general shape of the gamma-ray spectrum, which has remained relatively consistent from Pass~6 through Pass~7 Reprocessed, and even consistent with data from the COS-B and Compton Gamma-ray Observatory satellites \citep[see Figure 4 of][]{Casandjian2015}.

\subsection{Model Fitting Procedure}

The model was convolved with the instrument response functions and exposure using \texttt{gtsrcmaps}. The result is fit to the data with the \texttt{gtlike} tool, which compares the observed gamma-ray counts to the number of gamma-ray photons predicted by the model. When fitting Equation \ref{eq:model} to the gamma-ray data, only the normalizations are varied. First we identify point sources with flux $> 10^{12}$ ph \pers\ MeV$^{-1}$ cm$^{-2}$ and those within 5$^\circ$ of the center of the ROI. We fix the normalizations for the diffuse components of the model and fit the normalizations for the identified point sources with the \texttt{gtlike} optimizer DRMNFB. Finally we fix the point source parameters and fit the normalizations for the diffuse sources. We repeated this procedure at least one additional time with the optimizer NewMinuit. After the second iteration, the fit parameters do not change significantly. This procedure exploits the fact that diffuse sources and point sources are generally uncorrelated spatially. Note that this assumption breaks down for 3FGL confused sources, which were removed for this analysis.

The basic procedure for the likelihood analysis of gamma-ray data is described in \citet{Cash1979} and \citet{Mattox1996}. To evaluate source detection and model significance, we consider the likelihood ratio statistic, which we call $TS$ to match previous notation \citep[e.g.,][]{Ackermann2012c}, which is proportional to the difference of the log likelihoods of two different models after each model is maximized over its parameters:

\begin{equation}
    TS = -2\Big( \ln\mathcal{L}_A - \ln\mathcal{L}_B\Big),
\end{equation}

\noindent where $\mathcal{L}_A$ and $\mathcal{L}_B$ are the likelihoods for two models being compared (e. g., a source template versus a null). For each model, the combination of model parameters which maximizes that likelihood is found. The $TS$ is related to the significance of model $B$ over model $A$. The results do not change significantly if we consider a penalty on the addition of degrees of freedom, such as in the Bayesian information criterion \citep[BIC;][]{Schwarz1978}. This BIC adds a penalty in the form of $+\Delta\nu\ln(n)$, where $\Delta\nu$ is the difference in the number of degrees of freedom between the two models, and $n$ is the number of data points. Since models differ by at most four degrees of freedom, the $TS$ would change by at most $4\ln(30\times212^2) = 56$, where ($30\times212^2$) is the number of bins in the counts cube, and the difference in the number of degrees of freedom represents the \hi, \hii, \fermi\ bubbles, and isotropic components.

However, this likelihood ratio only compares the overall fits to each other. Two models may have similar likelihoods while having structured residuals remaining. Therefore, we also compute residual maps and $TS$ maps. A $TS$ map is calculated by rastering a point source across the region and refitting the model. Each pixel in the resulting map represents the improvement in the overall likelihood due to the inclusion of the additional point source. Large, coherent structures in the $TS$ map may represent unmodeled diffuse emission as opposed to unmodeled point sources.

\section{Results and Discussion}

\subsection{Extinction Compared with Diffuse Gamma-ray Counts from the ISM}

\begin{figure}
\centering
\includegraphics[width=\linewidth]{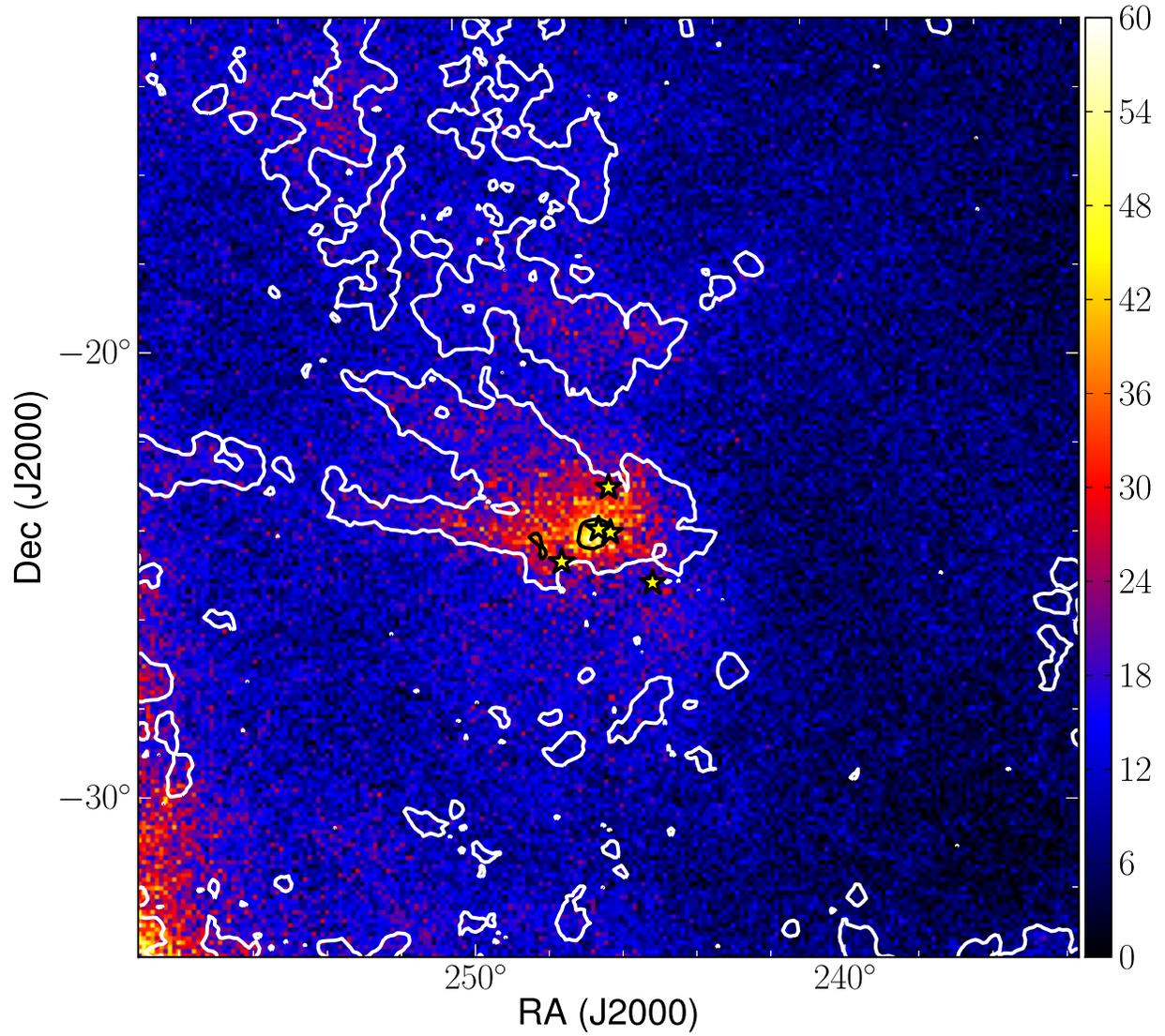}
\caption{Total gamma-ray counts minus the predicted counts from point sources. Thus, the image shows a prediction for the diffuse gamma-ray emission. Contours are as in Figure \ref{fig:CMAP}, and the stars indicate B stars near the ROMC (from east to west: 22 Sco, S1, \rOph, HD 147889, $\sigma$ Sco). The color scale is the same as Figure \ref{fig:CMAP}.}
\label{fig:CMAP_ISM}
\end{figure}

\begin{figure}
\includegraphics[width=\textwidth]{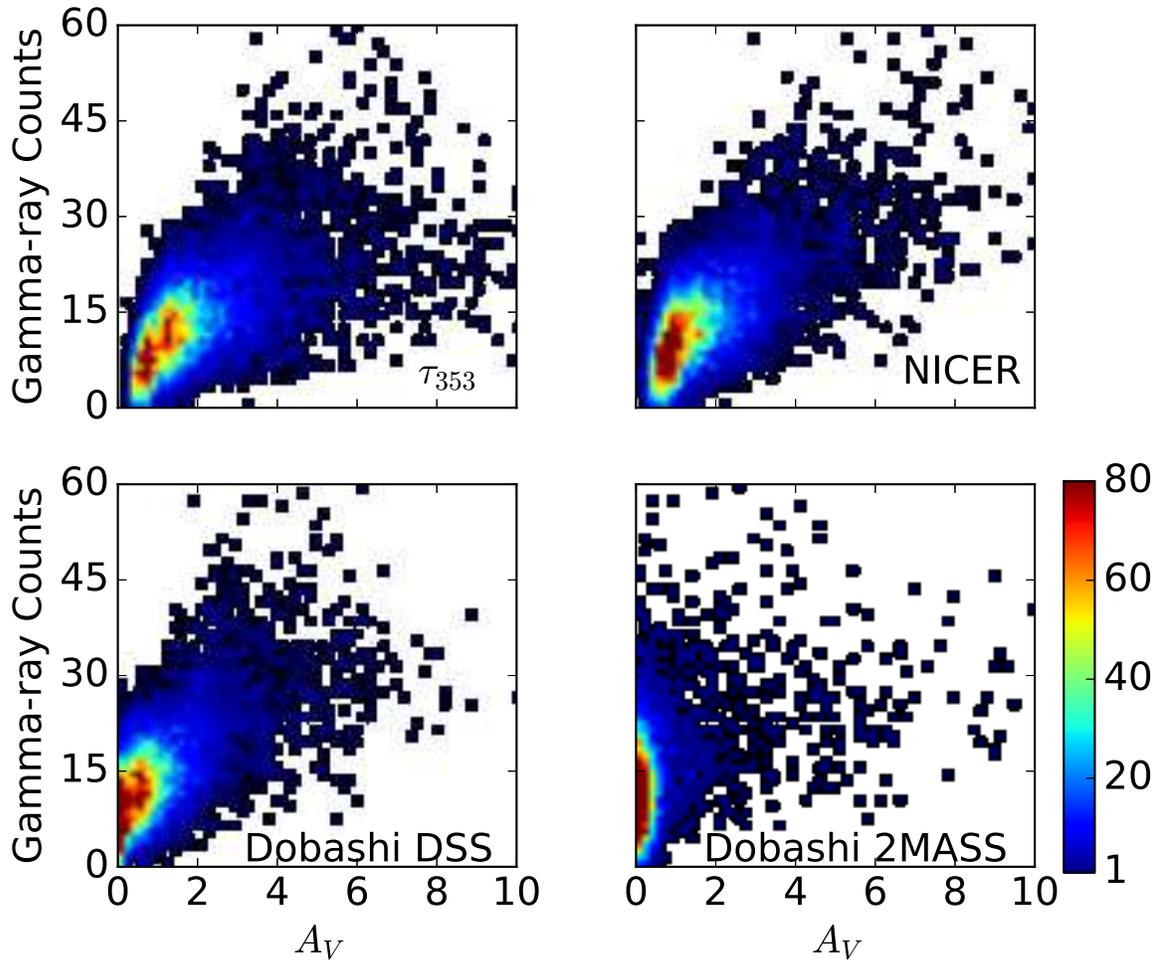}
\caption{Diffuse gamma-ray counts from the ISM (central $10\degr$ of Figure \ref{fig:CMAP_ISM}) versus \Av tracers, where the color scale represents the number of pixels. The extinction is cut off at 10 mag in order to highlight the relationship at low extinction.}
\label{fig:g-ray_comp}
\end{figure}

The total predicted counts for the 3FGL point sources, IC, and isotropic components may be subtracted from the counts map to render an estimation of the diffuse gamma-ray emission from the ISM alone (Figure \ref{fig:CMAP_ISM}). Figure \ref{fig:g-ray_comp} shows a pixel-by-pixel density plot of gamma-ray counts from the ISM versus $A_V$ for four gas templates. We include only the central 10$^\circ$ of the gamma-ray data to focus on the ROMC and minimize the influence of the IC component close to the Galactic center.    


Three of the tracers correlate reasonably well with the gamma-ray counts above $A_V \sim 0.5$ mag. The Dobashi DSS extinction does not reach above $A_V \sim 10$ mag, as expected due the limited optical star counts in highly extincted areas. The scatter is larger and the correlation weaker for Dobashi 2MASS extinctions.

The \TauEB\ and NICER extinctions correlate best with the gamma-ray counts, at least above $A_V \approx 1$ mag. There appears to be some nonlinearities below $\Av \approx 1$, close to the effective limit of both the \fermi\ and \Av\ data.

\begin{figure}
\centering
\includegraphics[width=\textwidth]{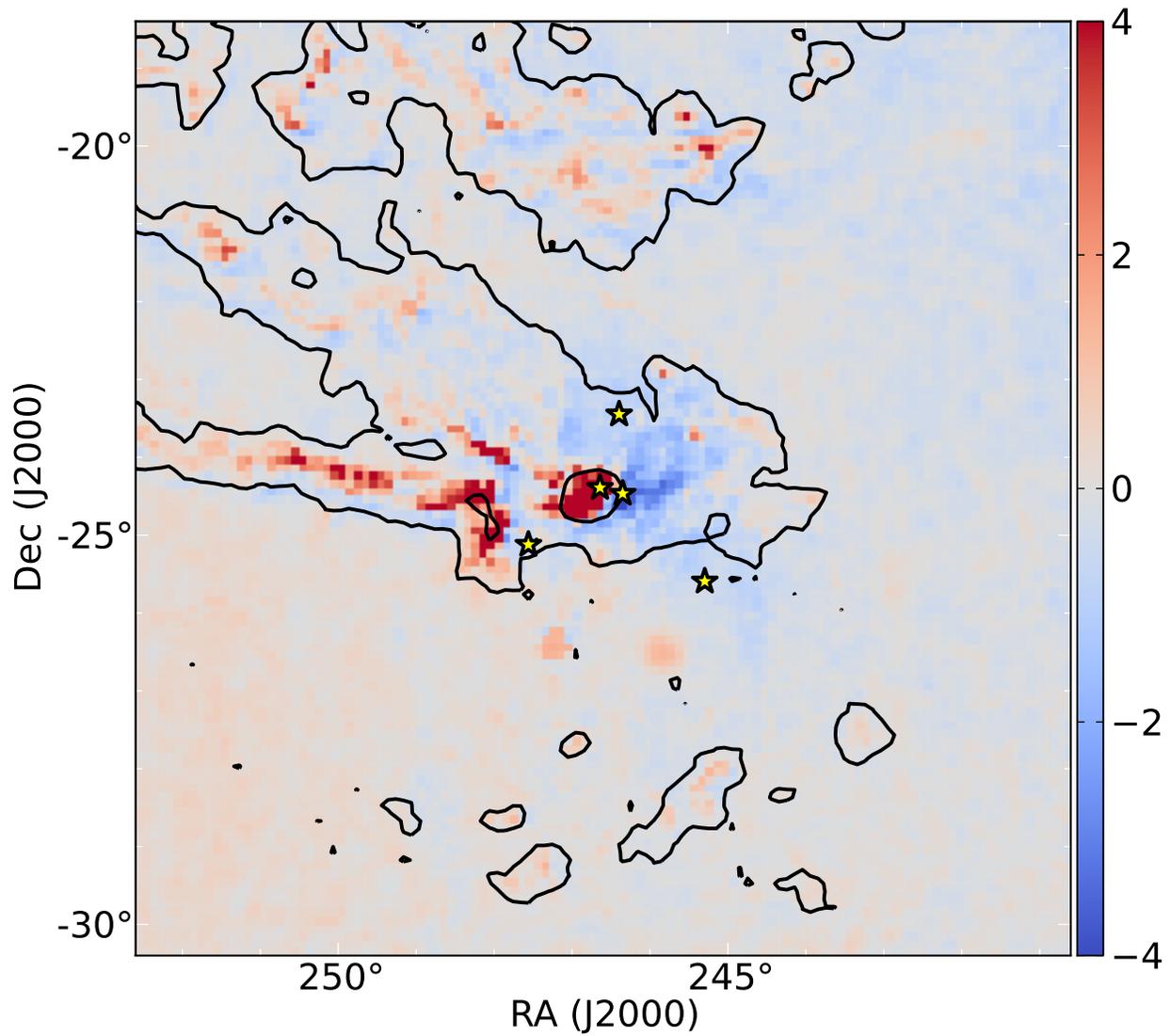}
\caption{Difference of extinctions $A_V(\TauEB)-A_V({\rm NICER})$. \Av(\TauEB) is multiplied by 0.89 in order to match the low extinction regions and highlight the differences. Contours represent \WCO\ emission and symbols are nearby B stars, as described in Figure \ref{fig:CMAP_ISM}.}
\label{fig:tau_nicer_ratio}
\end{figure}

\begin{figure}
\centering
\includegraphics[width=\linewidth]{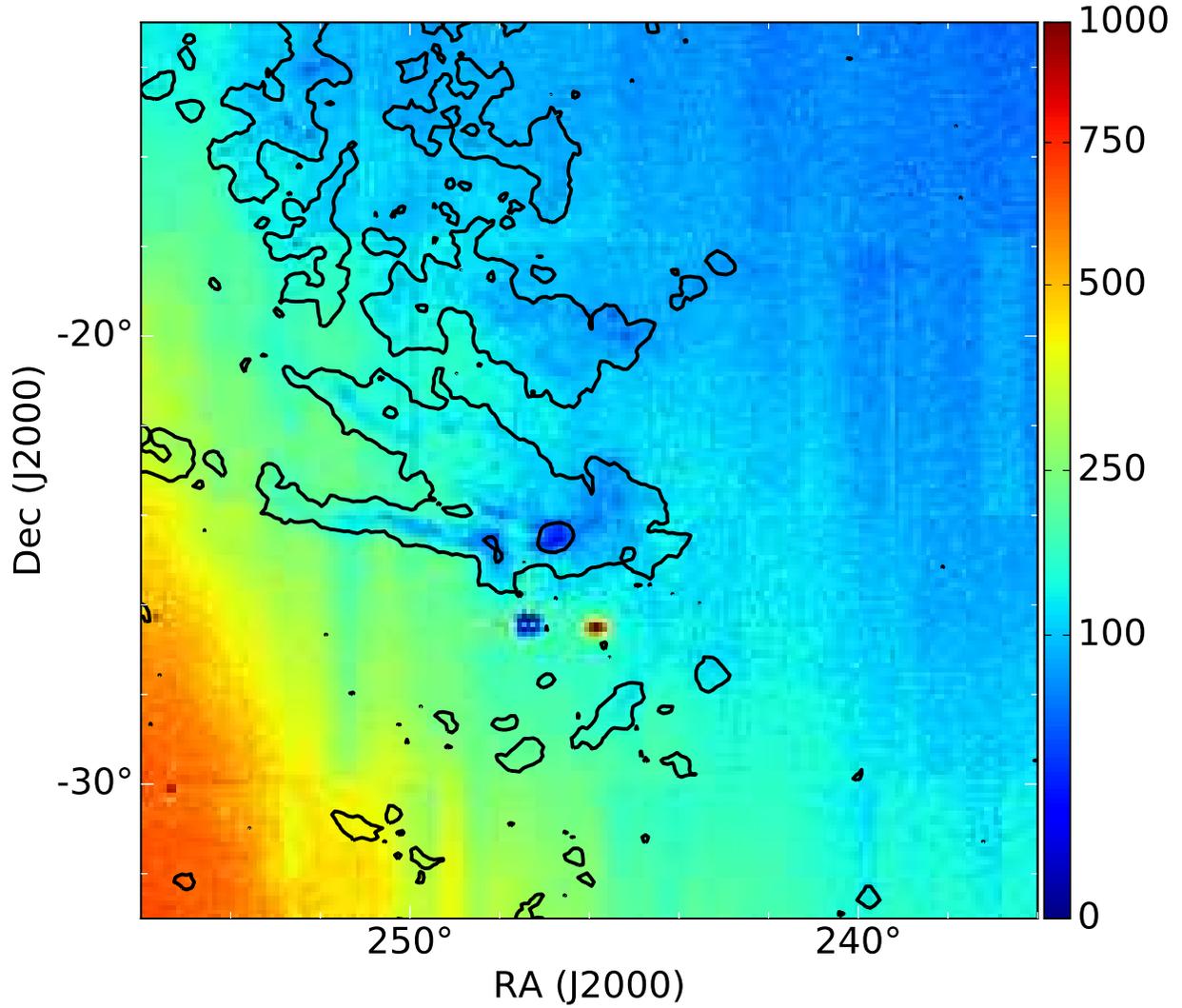}
\caption{The number of 2MASS stars per $6\arcmin$ pixel. The ROMC can be seen as a relative deficit of stars with $\le 100$ stars per pixel. The blue circular feature just south of the ROMC is masked due to the bright foreground star, Antares at (247\pdeg3, -26\pdeg4). The density enhancement west of Antares is the globular cluster, M4 at (245\pdeg9,-26\pdeg5). Contours are as in Figure \ref{fig:CMAP}.}
\label{fig:starcount}
\end{figure}

Despite their relatively good correlation, coherent and large-scale spatial differences between them are apparent in the maps. To highlight the differences between the two, we plot the difference $A_V(\TauEB)-A_V({\rm NICER})$ in Figure \ref{fig:tau_nicer_ratio}. Toward the highest integrated CO intensities, such as around L1688 and L1689, a higher $A_V$ is estimated from \TauEB\ than from NICER. These spots are coincident with 1.1 mm continuum sources \citep{Young2006}, and may represent a bias in the \TauEB\ maps.  Some regions where the \TauEB\ extinction is lower, such as around the star $\rho$ Oph north of L1688, exhibit both elevated dust temperatures and lower dust emissivity power law indices ($\beta$), indicative of varying dust properties. These spots, however, may also suffer from a sampling bias in the NICER procedure. Sub-pixel scale structure, particularly toward highly extincted areas, will preferentially allow starlight to pass through relatively diffuse gaps in the cloud, thus biasing the 2MASS extinctions downward. A stellar density map (Figure \ref{fig:starcount}) confirms that these areas show marked decrements. \citet{Lombardi2009} developed the NICEST algorithm to quantitatively address this issue, which we comment on further in a later subsection.

\subsection{Model Component Contributions and Significance}

\begin{figure}
\begin{minipage}{0.5\textwidth}
\centering
\includegraphics[width=\linewidth]{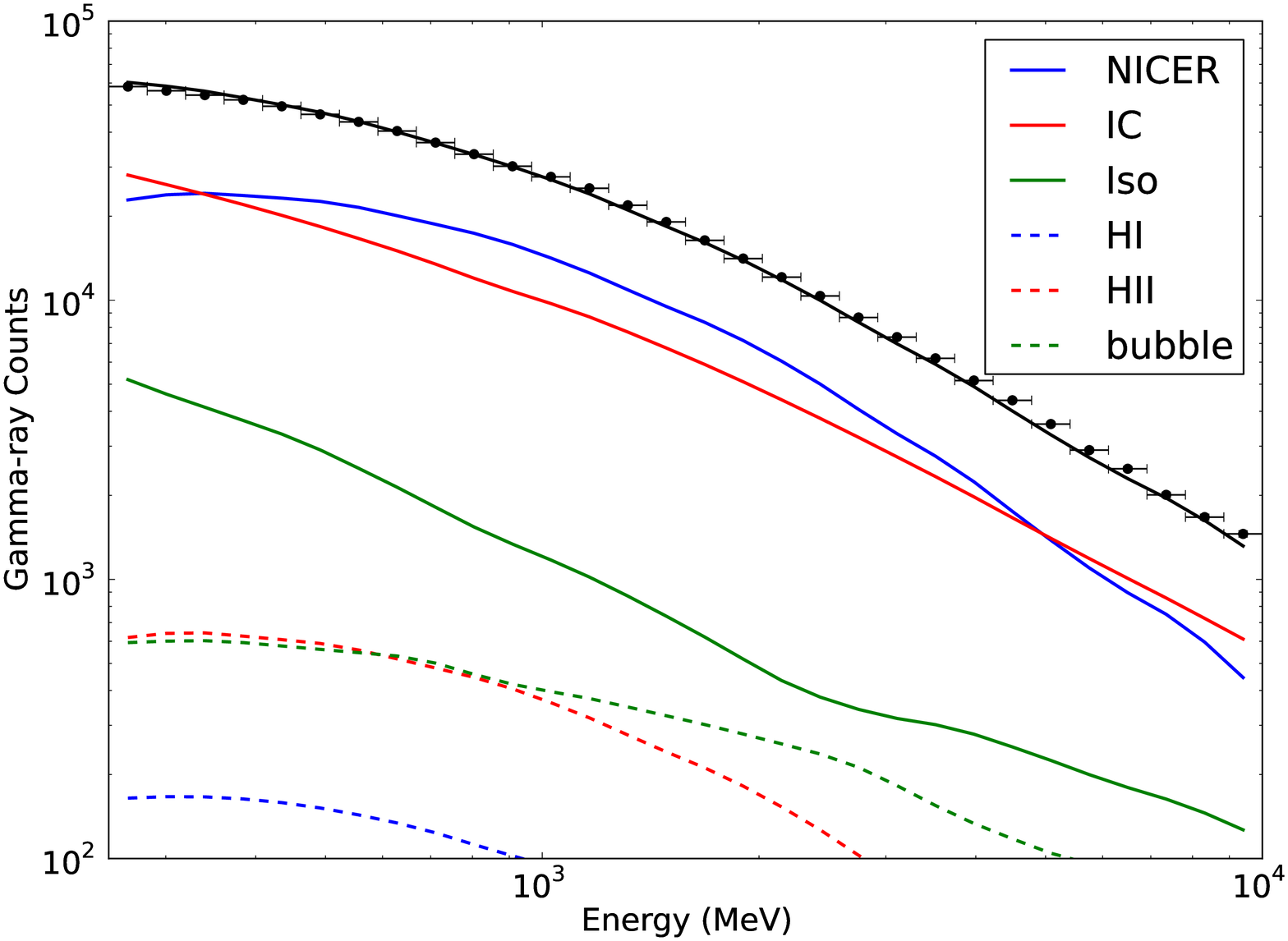}
\end{minipage}
\begin{minipage}{0.5\textwidth}
\centering
\includegraphics[width=\linewidth]{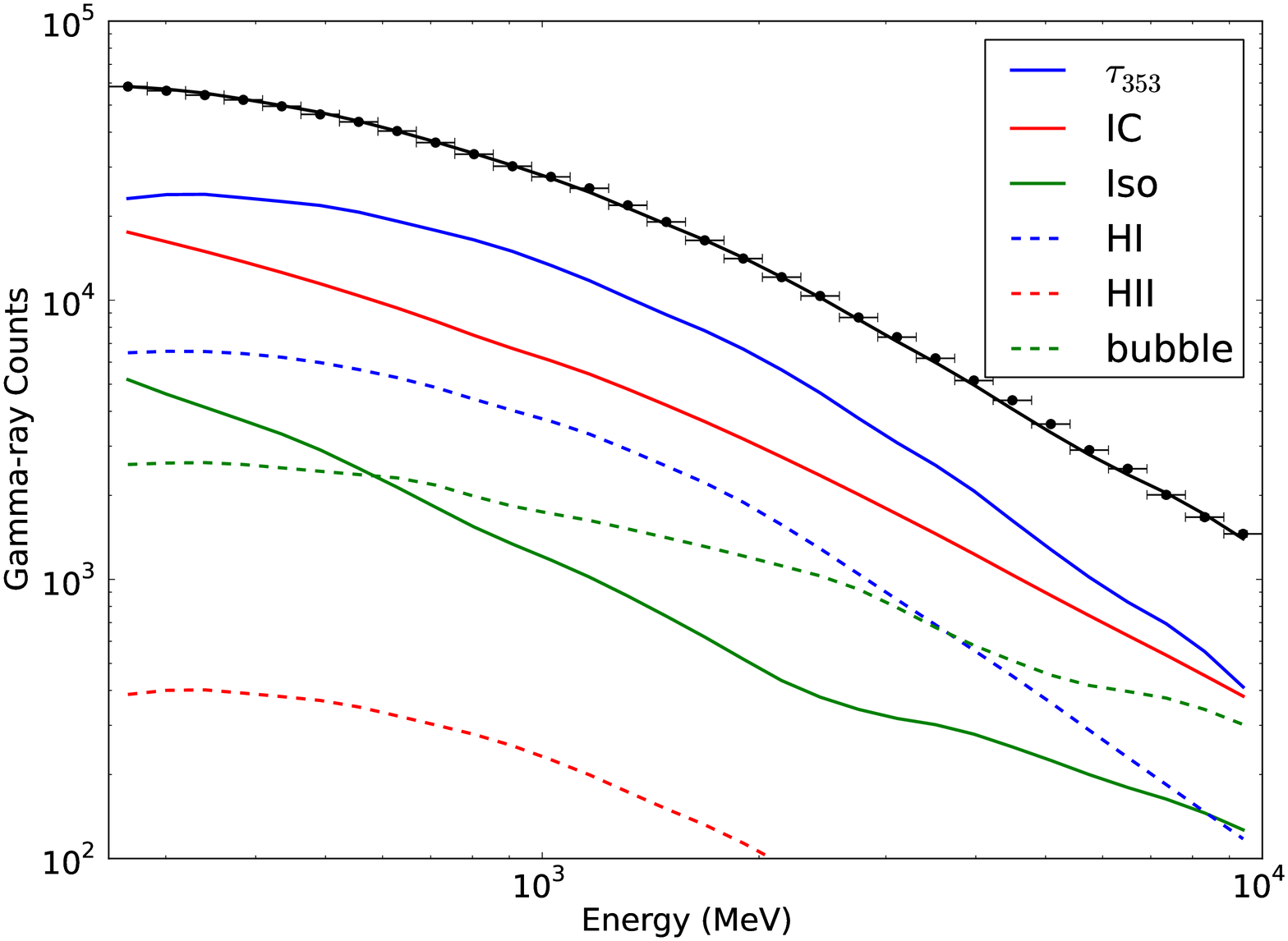}
\end{minipage}
\caption{Observed versus predicted gamma-ray counts for the model and for diffuse components of the model for NICER (left) and \TauEB\ (right). The black dots are the data and the black line is the sum of all model components. The \Av\ tracer and IC components dominate the integrated gamma-ray emission across the entire ROI.}
\label{fig:count_spec}
\end{figure}

Over the whole ROI, the gas components ({ ${A_V}$} tracers) dominate the gamma-ray spectrum.  While the \Av templates contribute { to the gamma-ray emission} more than the additional 21~cm template alone by over an order of magnitude, the \hi\ component is still detected at a statistically significant level. This result suggests that the \Av\ map does not completely trace { the ISM}. \hii\ also contributes less than an order of magnitude towards the total predicted gamma-ray counts, seen for both the NICER and the \TauEB\ tracers in Fig \ref{fig:count_spec}. These spectra are integrated across the entire region. Both the \Av\ and IC components dominate the counts spectrum, while the isotropic component contributes an order of magnitude less. Both \hi\ and \hii\ components contribute significantly less than the \Av\ component. They are also spatially uncorrelated with the ROMC. Thus, we establish that \Av\ traces most of the diffuse gamma-ray emission from the ISM. However, \hi\ and \hii\ are still required when integrating across the entire ROI.

Table \ref{tab:results} lists the results from the gamma-ray likelihood analysis for the diffuse emission. The normalization for the IC contribution is notably high (greater than unity) in all cases. Although this may also indicate a poor choice of GALPROP inputs, it is unimportant to our discussion of the cloud structure given the lack of small-scale structure in the IC emission over the ROI.

The $TS$ value listed in Table \ref{tab:results} compares each model to the \fermi\ diffuse model, where the baseline model analyzes the ULTRACLEAN VETO data class and includes: the IC from GALPROP galdef file ``54 77Xvarh7S", the isotropic component normalization fixed at unity, the  \fermi\ bubbles, \hi, and \hii. Again, none of the models include the two confused point sources coincident with L1688. A positive $TS$ value represents a better fit considering the entire ROI. It is important to note that for extended emission and large fields, judging the fit from this $TS$ value alone can be misleading. In trying to determine which gas template recovers the cloud emission best, we examine not only the $TS$ value across the ROI, but also $TS$ maps and residual maps, described below. Only the model with \TauEB\ has a higher likelihood across the ROI than the \fermi\ diffuse model, with ${TS=89}$. The higher $TS$ stems almost entirely from accounting for the additional gamma-ray emission in L1688. The Galactic center excess, which is accounted for in the \fermi\ diffuse model, is another significant contributor to the likelihood differences when considering the entire ROI. 

To quantify the importance of the various diffuse model components, we alternately modify each component in the model or change the gamma-ray data cut. We report the maximum percentage difference from the baseline model in Table \ref{tab:sys_uncertainty}. The dominant systematic uncertainty occurs when \hi\ is not included in the model. The IC contribution in many cases is competitive with the \Av\ component, which is a consequence of its near isotropy over the ROI -- it contributes to every pixel -- and the strong expected IC signal towards the Galactic center. This is also seen in the predicted gamma-ray counts spectra in Figure \ref{fig:count_spec}. The other model components, such as the isotropic background and \fermi\ bubbles, contribute very little to the emission. In fact, the reduced photon statistics used to fit these components can make it difficult to spatially or spectrally distinguish them. The covariance matrix resulting from \texttt{gtlike} fits also indicates significant correlations between these weaker components.

%

\subsection{Residual and \emph{TS} Maps}
\label{ssec:result_resid}
The counts map is subtracted from the model generated from \texttt{gtmodel}, which which convolves the model with the exposure, point spread function, and instrument response function, to generate a residual map. Regions where the model underestimates the emission, then, result in a positive residual. Significance maps are generated by

\begin{equation}
\frac{N_{data} - N_{model}}{\sqrt{N_{model}}}.
\end{equation}

\begin{figure}
\includegraphics[width=\textwidth]{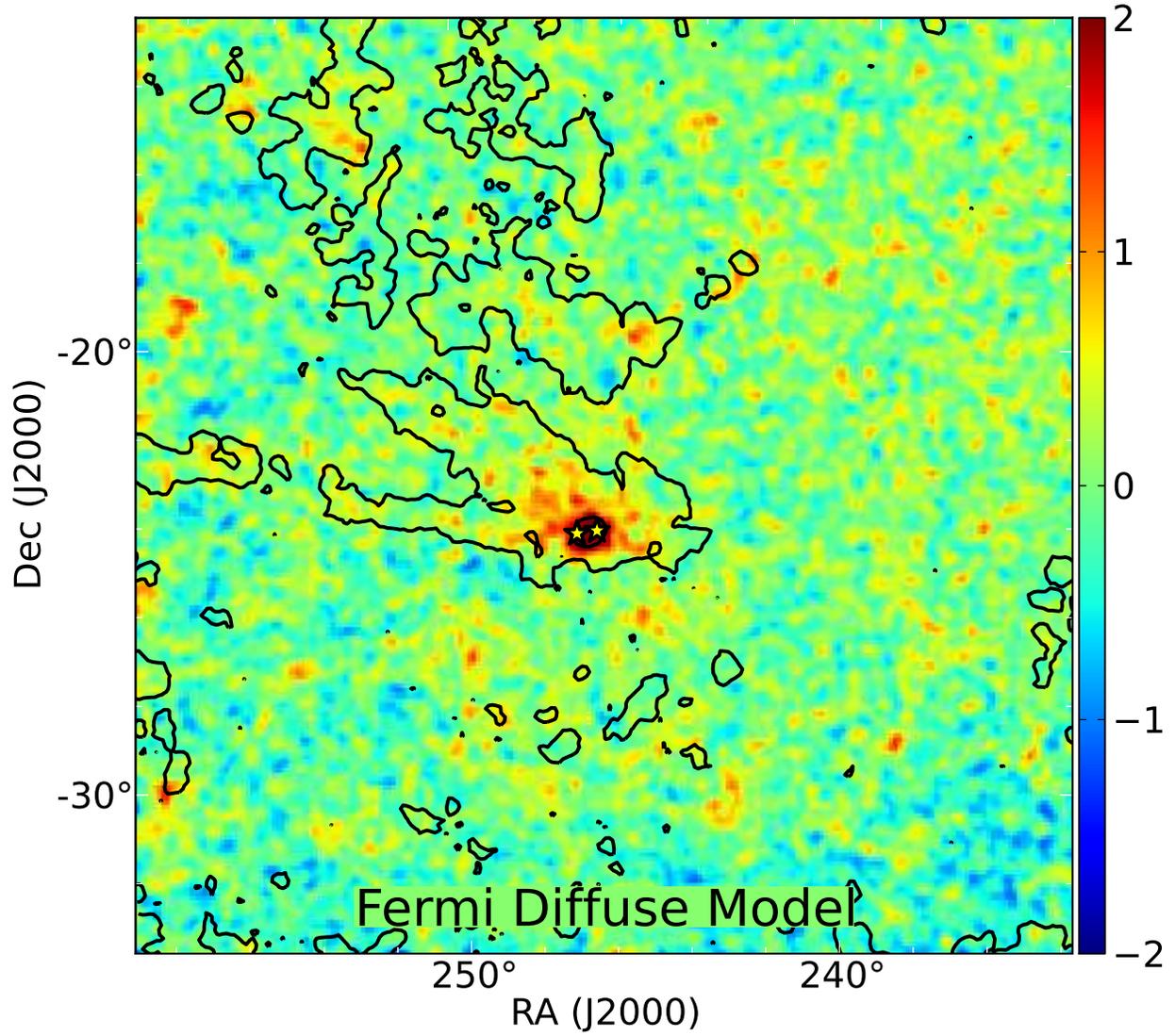}
\caption{Residual significance map for the \fermi\ diffuse model smoothed with a 0\pdeg2 Gaussian kernel. The yellow stars mark the positions of the two confused point sources in the 3FGL, and contours are as in Figure \ref{fig:CMAP}.}
\label{fig:galdiff_resid}
\end{figure}

\noindent
The \fermi\ Galactic diffuse model, omitting the confused point sources 3FGL J1628.2-2431c and 3FGL J1626.2-2428c, produces a significant gamma-ray residual around L1688 (Figure \ref{fig:galdiff_resid}). When included, these point sources have $TS$ values of 305 and 189, and photon indices of $-3.0$ and $-2.2$ for the eastern and western sources, respectively. Such $TS$ values indicate their significance. Their photon indices and lack of variability are typical for ISM emission. It is unclear why the \fermi\ diffuse model results in such a substantial residual at the CO emission peak of the ROMC, when it incorporates the \citet{Dame2001} CO map.

\begin{figure}
\includegraphics[width=\textwidth]{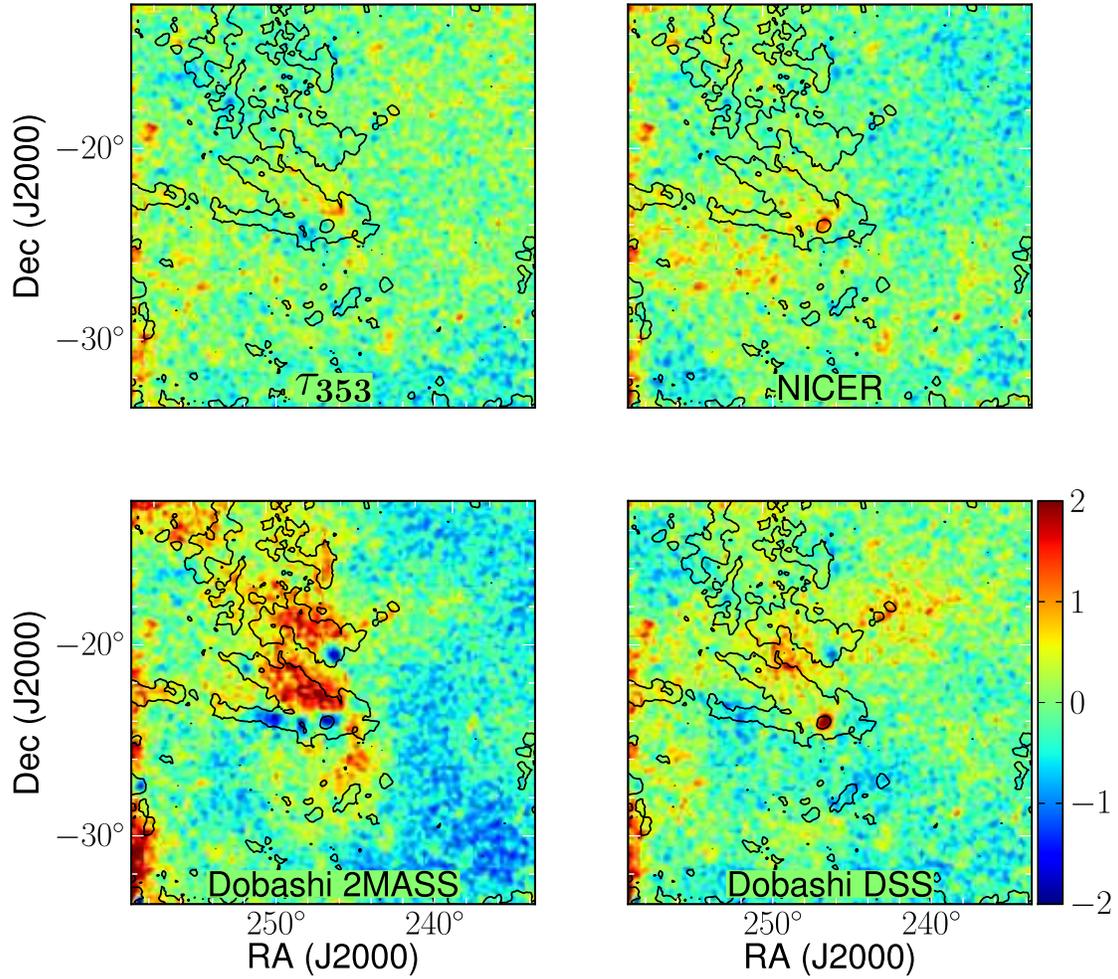}
\caption{Residual significance maps for \TauEB\ (top left), NICER (top right), Dobashi 2MASS (bottom left), and Dobashi DSS (bottom right) smoothed with 0\pdeg2 Gaussian kernel. Contours are as in Figure \ref{fig:CMAP}.}
\label{fig:resids_all}
\end{figure}

Figure \ref{fig:resids_all} shows that \TauEB\ and NICER perform significantly better -- that is, have fewer and smaller spatially coherent residuals -- than either Dobashi map, and better than the \fermi\ diffuse model near L1688. The NICER template underestimates the gamma-ray emission in the densest part of the ROMC, inside the western $\WCO=50$ K \kms\ contour. The \TauEB\ model overestimates the emission within the eastern $\WCO=50$ K \kms\ contour near the L1689 cluster and the B star 22 Sco{, and underestimates the gamma-ray emission} just north of L1688 around the B star $\rho$ Ophiuchi. Both of the Dobashi maps show significant, structured, positive and negative residuals across the entire ROI. These failures occur in both high and low density regions. The Galactic center excess is also seen in the southeast corner. While this signal is accounted for in the \fermi\ Galactic diffuse model, it is not included in our gas templates. It is thus a persistent residual in this study, but also physically distinct from the emission of the ROMC. \TauEB\ reproduces the gamma-ray emission from the Galactic center region better than NICER or NICEST, which gives \TauEB\ the highest $TS$ value in Table \ref{tab:results}

To further quantify the significance of the spatial residuals, we show $TS$ maps for the \fermi\ diffuse model, NICER, and \TauEB\ in Figure \ref{fig:TSmaps}. The $TS$ maps focus on the immediate region around L1688. The \texttt{gttsmap} routine places a point source at each pixel and evaluates its $TS$ compared to the null. While the $TS$ map can achieve higher resolution than the residual maps, the improvement of adding a point source can be assessed only if the original model underestimates the gamma-ray counts. The \TauEB\ and NICER models used to make the $TS$ maps include the \hi\ component, but not the \fermi\ bubbles. Because the \fermi\ bubbles template is isotropic across the ROI, it does not affect the structure of the $TS$ maps. 

The $TS$ map for NICEST shows the least structure, and thus it does not significantly underestimate the gamma-ray emission within the ROMC. The $TS$ map for the \fermi\ diffuse model, \TauEB, and NICER contain a regions of very high $TS$ values near L1688, similar to its residual map. The \TauEB\ model fails to the north of L1688 around the B star system $\rho$ Ophiuchi (note again that the overestimate near L1689 is not captured in the $TS$ map). The $TS$ map for the NICER model primarily indicates a failure in the densest regions of the ROMC.

\subsection{Cosmic Ray Sources in the ROMC}

The NICER model appears to be the best overall fit for the ROMC based on the spatial residuals and integrated values of the $TS$ maps. The lone residual appears in the densest part of the ROMC near L1688. An exciting possibility is that the young star-forming region could be a source of CRs, although a more likely explanation for the residual gamma-ray signal is improper modeling of the ISM. As mentioned before, there is a known bias for photometric extinction methods to underestimate the column density towards dense regions. Indeed, the underestimate at L1688 is most severe for the Dobashi DSS template, which is expected given the quicker saturation of the optical extinction. Using the Dobashi DSS template, we insert a point source with a power law spectrum into the model at the location of L1688, which returns a $TS$ value of 460, a photon index of $-2.48$, and flux of $2.38 \times 10^{-8}$ photons cm$^{-2}$ s$^{-1}$. We would expect a harder photon index from freshly accelerated CRs, but this putative source merely resembles the two confused 3FGL sources and typical ISM emission.


As mentioned before, small-scale cloud structure allows starlight to pass through relatively diffuse gaps in the clouds, thus causing a downward bias in extinction from the NICER and Dobashi methods. \citet{Lombardi2009} found this effect to be most severe in the thickest regions of a cloud where there is a distinct decrease in star counts (Figure \ref{fig:starcount}). In response, \citet{Lombardi2009} developed the NICEST algorithm, which compensates for this bias using the star counts themselves.

\begin{figure}
\includegraphics[width=\textwidth]{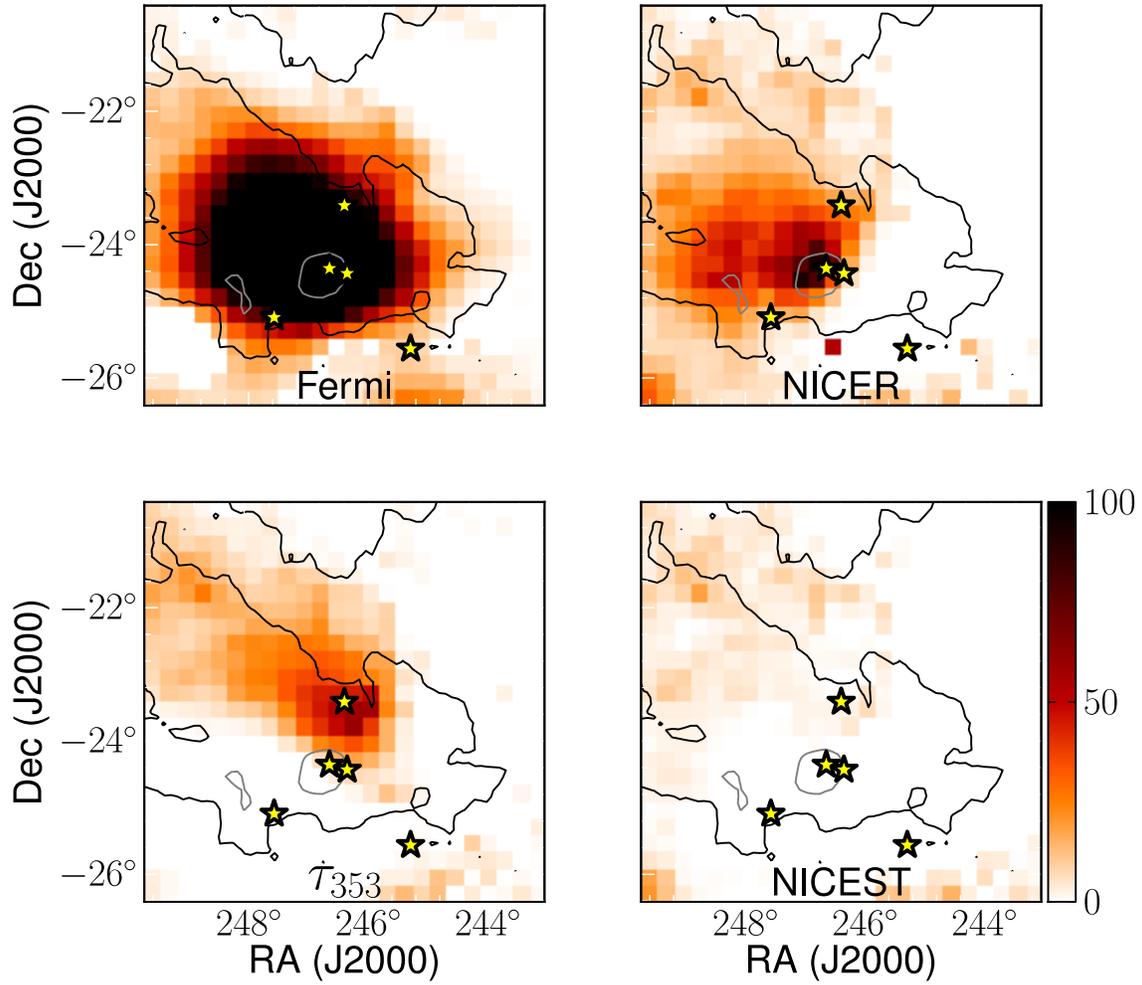}
\caption{TS maps for the \fermi\ Galactic diffuse model (top left), the  NICER extinction model (top right), the \TauEB\ model (bottom left), and the NICEST extinction model (bottom right). All images have the same color scale. Contours and symbols are as in Figure \ref{fig:CMAP_ISM}.}
\label{fig:TSmaps}
\end{figure}

To check for this bias, we model the gamma-ray observations from the ROMC with the NICEST map from \citet{Juvela2016}. The model excludes the \fermi\ bubbles. As shown in Figure \ref{fig:TSmaps}, there are no significant gamma-ray residuals in the ROMC; the $TS$ map shows no missing sources. Thus L1688 can be well modeled simply with diffuse emission templates. This result implies that L1688 is not an important local CR acceleration site. The NICEST extinction map yields the best recovery of the gamma-ray data of all the models tested in this study.

\begin{figure}
\centering
\includegraphics[width=\textwidth]{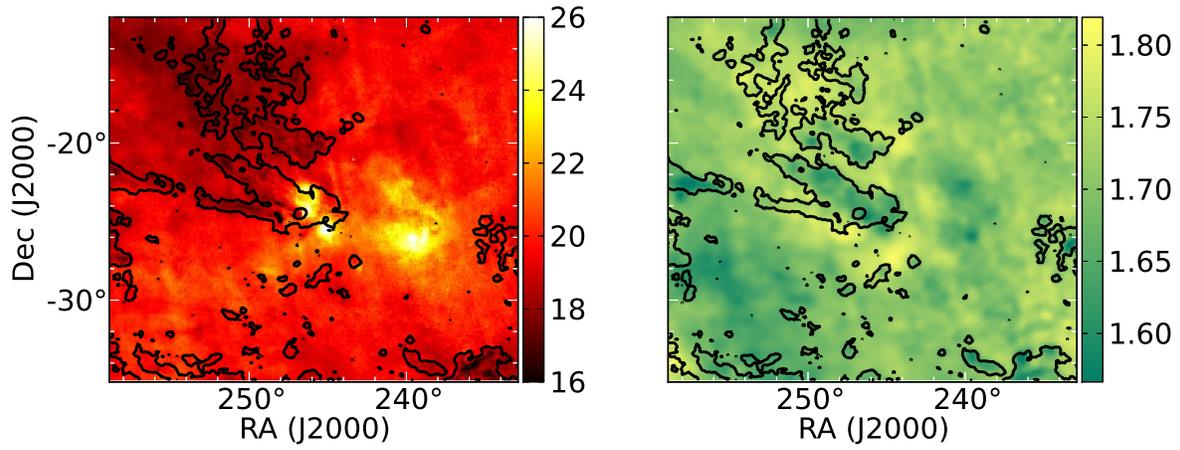}
\caption{(Left) Dust temperature in K and (right) dust emissivity power law index $\beta$ around the ROMC as determined by \planck. Contours are as in Figure \ref{fig:CMAP}.}
\label{fig:dust_T_beta}
\end{figure}

The \planck\ \TauEB\ model underestimates the gamma-ray emission toward the $\rho$ Ophiuchi star system, but overestimates the emission toward L1689 and 22 Sco. It may be that embedded hot stars alter the optical properties of the dust, causing a systematic offset in \TauEB. The region of high $TS$ for the \TauEB\ model in Figure \ref{fig:TSmaps} is coincident with elevated dust temperatures (Figure \ref{fig:dust_T_beta}). The locations of other embedded young stars, especially near L1688, similarly correspond to hot dust and variations in $\beta$ (Figure \ref{fig:dust_T_beta}). So while \TauEB, including \hi, seems to be a reasonable model in diffuse regions across the ROI, towards dense regions -- especially those with embedded hot stars -- the degeneracy between dust temperature and the dust spectral index may hamper its universality as a column density tracer.

\section{Conclusions}

We modeled the gamma-ray emission of the ROMC with different \Av\ tracers in order to determine which most accurately estimates column density in molecular clouds. Infrared dust optical depth at 353 GHz correlates well with the gamma-ray emission over the entire $20\degr\times20\degr$ ROI, and even toward L1688, but may suffer some small systematic effects due to varying dust properties around embedded hot stars. Without correcting for the downward extinction bias caused by small-scale structure in very dense regions, the NICER method clearly underestimates the gamma-ray emission, and thus the column density, towards the dense core around L1688. The NICEST method attempts to correct for biases due to high extinction. As a result, NICEST most successively estimates the total gas column density as traced by the gamma-ray emission.

NICER, NICEST, and \TauEB\ all suffer from noise constraints at low extinction, and thus cannot trace the ISM in low density regions. For NICER and NICEST, this is due to the intrinsic spread in stellar colors. \TauEB\ suffers from uncertainties in modeling the infrared spectrum. The gamma-ray fit is improved significantly by including an \hi\ component. This result verifies the claim that a combination of \hi\ and far IR dust emission better traces \Av\ than either alone \citep{Peek2013}.

We are able to recover all of the diffuse gamma-ray emission in the ROMC with templates based on gas or dust emission and/or extinction measurements. We find no evidence for any additional sources of CRs. Specifically, the young star cluster L1689 does not appear to be a significant CR acceleration site.


\acknowledgments

The authors thank the \textit{Fermi} LAT team for their support. This work was supported in part by the NASA New York Space Grant Consortium based at Cornell University (\# NNX10AI94H) and by the NSF (AST-1153335). Support for AT was provided by Undergraduate Research Fellowships and the Raab Presidential Fellowship at CUNY Hunter College. This research has made use of the NASA/IPAC Infrared Science Archive, which is operated by the Jet Propulsion Laboratory, California Institute of Technology, under contract with the National Aeronautics and Space Administration.

{\it Facilities:} \facility{\textit{Fermi} (LAT)}

\clearpage

\clearpage

\clearpage

\clearpage

\clearpage

\clearpage

\clearpage

\clearpage

\clearpage

\clearpage

\clearpage

%


\clearpage
\begin{deluxetable}{l c c c c c c c}
\tablecolumns{8}
\tablewidth{0pt}
\rotate
\tablecaption{Results for extinction tracers}
\tablehead{
\colhead{} 				&
\colhead{} 				&
\multicolumn{6}{c}{Normalization}\\ 

\cline{3-8} \\

\colhead{MODEL}			     	&
\colhead{$TS$\tablenotemark{a}}&
\colhead{\Av\ Tracer}			&
\colhead{\hi}					&
\colhead{IC}					&
\colhead{Isotropic}				&
\colhead{Bubbles}				&
\colhead{\hii}
}

\startdata

\TauEB & 89 & $1.06\pm0.01$ & $0.29\pm0.02$ & $2.34\pm0.03$ & 1 & $1.30\pm0.05$ & $(3.91\pm0.67)\ee{-4}$ \cr

NICEST & -691 & $0.44\pm0.01$ & $0.13\pm0.02$ & $3.37\pm0.03$ & 1 & $0.80\pm0.05$ & $0.07\pm0.01$ \cr

NICER & -1609 & $0.43\pm0.01$ & \nodata & $3.75\pm0.02$ & 1 & $0.30\pm0.05$ & $0.06\pm0.01$ \cr

Dobashi DSS & -3265 & $0.55\pm0.01$ & $1.10\pm0.02$ & $2.15\pm0.03$ & 1 & $1.57\pm0.06$ & $0.01\pm0.01$ \cr

Dobashi 2MASS & -12556 & $1.11\pm0.01$ & $1.34\pm0.01$ & $2.56\pm0.03$ & 1 & \nodata & \nodata \cr \hline
\enddata
\tablecomments{The various dust tracers compared to the \fermi\ diffuse model and associated component normalizations ordered by $TS$.}\label{tab:results}
\tablenotetext{a}{Model likelihood for the entire ROI compared to the $-\ln\mathcal{L}$ of the \fermi\ diffuse model after removing the two point sources associated with the ROMC.}
\end{deluxetable}

\clearpage

%
%
%
\begin{deluxetable}{l c c c c c c c}
\tablecolumns{8}
\tablewidth{0pt}
\tablecaption{Systematic Uncertainties}
\tablehead{

\colhead{Model}			&
\colhead{Data Cut}			&
\colhead{IC Model}			&
\colhead{Iso free}			&
\colhead{Bubbles}			&
\colhead{\hi}				&
\colhead{\hii}				&
\colhead{Total}			\\
\colhead{}				&
\colhead{(1)}				&
\colhead{(2)}				&
\colhead{(3)}				&
\colhead{(4)}				&
\colhead{(5)}				&
\colhead{(6)}				&
\colhead{(7)}			
}
\startdata
$\tau_{353}$		& 3.5 	& 0.2 & 0.4 & 2.2 & 15.8 	& 0.1 & 16.3\cr
NICER			& 8.8 	& 6.5 & 2.1 & 2.7 & 6.3 	& 0.8 & 13.1 \cr
NICEST			& 5.9 	& 3.9 & 0.1 & 0.8 & 9.5 	& 0.1 & 11.9  \cr
Dobashi DSS		& 7.9 	& 0.6 & 0.0 & 2.9 & 16.9 	& 0.1 & 18.9 \cr
Dobashi 2MASS	& 11.9	& 5.7 & 2.6 & 4.9 & 57 		& 1.7 & 58.8 
\enddata
\tablecomments{Absolute value of the differences in \Av\ normalization among all models fitted for each \Av, as a percent difference from the baseline model. Column (1) Data Cut compares the  SOURCE versus ULTRACLEAN VETO data cuts; Column (2) IC Model compares the two IC templates calculated from GALPROP; Column (3) Iso Free compares whether the isotropic component was fixed at one or set as a free parameter; Columns (4) -- (6) compare the baseline model to those that omit the \fermi\ Bubbles, \hi, or \hii\ components, respectively; Column (7) Total is calculated by adding the other columns in quadrature.}\label{tab:sys_uncertainty}
\end{deluxetable}

\end{document}